\documentstyle[epsfig]{aipproc}
\newcommand{\beq}{\begin{equation}}
\newcommand{\eeq}{\end{equation}}
\def\beqa{\begin{eqnarray}}
\def\eeqa{\end{eqnarray}}
\def\lap{\lower.5ex\hbox{$\; \buildrel < \over \sim \;$}}
\def\gap{\lower.5ex\hbox{$\; \buildrel > \over \sim \;$}}

\begin{document}
\title{The quantum cosmology debate}
\author{Alexander Vilenkin}
\address{Institute of Cosmology,\\Department of Physics,
Tufts University, Medford MA 02155}
\maketitle

\begin{abstract}

I review the tunneling, Hartle-Hawking and Linde proposals for the
wave function of the universe and comment on the recent work on
quantum creation of open universes.

\end{abstract}

\section*{Introduction}

In quantum cosmology the whole universe is treated
quantum-mechanically and is described by a wave function rather than
by a classical space-time.  This quantum approach to cosmology may
help us avoid the cosmological singularity problem and 
understand what determined the initial state of the universe.

The wave function of the universe $\psi$ satisfies the Wheeler-DeWitt
equation
\beq
{\cal H}\psi=0
\eeq
which is analogous to the Schrodinger equation in ordinary quantum
mechanics.  To solve this equation, one has to specify some boundary
conditions for $\psi$.  In quantum mechanics, the boundary conditions
are determined by the physical setup external to the system.  But
since there is nothing external to the universe, it appears that
boundary conditions for the wave function of the universe should be
postulated as an independent physical law.  The possible form of this law has
been debated for some 15 years and is one of the central points of our
debate here.  

There are at least three proposals on the table:  
the Hartle-Hawking wave function \cite{HH}, the Linde wave function
\cite{L1}, and the tunneling wave function \cite{V84,V86}.  In
this talk I will first review the tunneling wave function.  Then,
since this is supposed to be a debate, I will attack my opponents.
And finally, I will comment on the recent work on quantum creation of
open universes.

\section*{The tunneling wave function}

To introduce the tunneling wave function, let us consider a very
simple model of a closed Robertson-Walker universe filled with a
vacuum of constant energy density $\rho_v$ and some radiation.  The
total energy density of the universe is given by
\beq
\rho=\rho_v+\epsilon/a^4,
\label{1}
\eeq
where $a$ is the scale factor and $\epsilon$ is a constant
characterizing the amount of radiation.  
The evolution equation for $a$ can be
written as
\beq
p^2+a^2-a^4/a_0^2=\epsilon.
\label{2}
\eeq
Here, $p=-a{\dot a}$ is the momentum conjugate to $a$ and 
$a_0=(3/4)\rho_v^{-1/2}$.

\begin{figure}[b!] 
\centerline{\epsfig{file=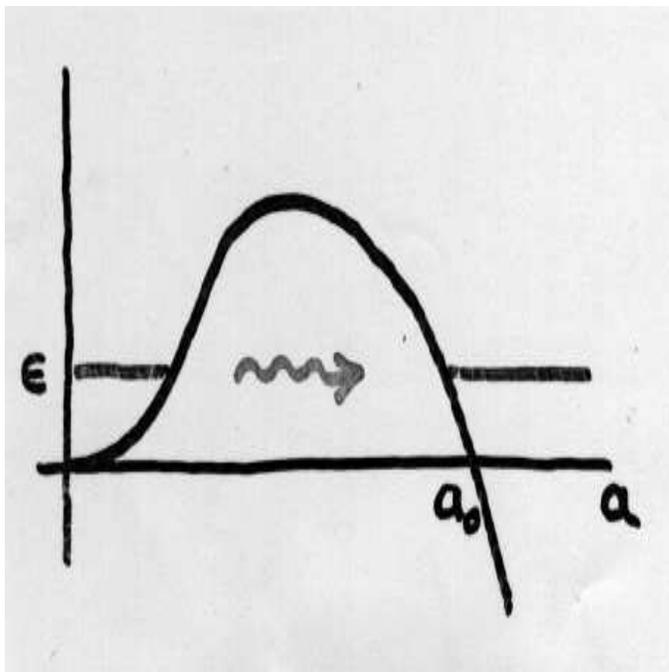,height=3.5in,width=3.5in}}
\vspace{10pt}
\caption{The potential for the scale factor in Eq.(\ref{2}).  Instead
of recollapsing, the universe can tunnel through the potential
barrier to the regime of unbounded expansion.}
\label{fig1}
\end{figure}
 
Eq.(\ref{2}) is identical to that for a ``particle'' of energy
$\epsilon$ moving in a potential $U(a)=a^2-a^4/a_0^2$.  For
sufficiently small $\epsilon$, there are two types of classical trajectories.
The universe can start at $a=0$, expand to a maximum radius $a_1$ and
then recollapse.  Alternatively, it can contract from 
infinite size, bounce at a minimum radius $a_2$ and then re-expand
(see Fig. 1).  But in quantum cosmology there is yet another
possibility.  Instead of recollapsing, the universe can
tunnel through the potential barrier to the regime of unbounded
expansion.  The semiclassical tunneling probability can be estimated
as
\beq
{\cal P}\sim\exp\left(-2\int_{a_1}^{a_2} |p(a)|da\right).
\eeq
It is interesting that this probability does not vanish in the limit
of $\epsilon\to 0$, when there is no radiation and 
the size of the initial universe shrinks to
zero.  We then have tunneling from {\it nothing} to a closed universe
of a finite radius $a_0$; the corresponding probability is
\beq
{\cal P}\sim\exp\left(-2\int_0^{a_0} |p(a)|da\right)
=\exp\left(-{3\over{8\rho_v}}\right).
\eeq
The tunneling approach to quantum cosmology assumes that 
our universe originated in a tunneling event of this kind.  Once it
nucleates, the universe immediately begins a de Sitter inflationary
expansion.

The Wheeler-DeWitt equation for our simple model can be obtained by
replacing the momentum $p$ in (\ref{2}) by a differential operator,
$p\to -id/d a$,
\beq
\left({d^2\over{da^2}}-a^2+{a^4\over{a_0^2}}\right)\psi(a)=0.
\label{WDW}
\eeq
This equation has outgoing and ingoing wave solutions corresponding to
expanding and contracting universes in the classically allowed range
$a>a_0$ and exponentially growing and decaying solutions in the
classically forbidden range $0<a<a_0$.
The boundary condition that selects the tunneling wave function
requires that $\psi$ should include only an outgouing wave at
$a\to\infty$ (see Fig.2). The under-barrier wave function is then a linear
combination of the growing and decaying solutions.  The two solutions
have comparable magnitudes near the classical turning point, $a=a_0$,
but the decaying solution dominates in the rest of the under-barrier
region.

\begin{figure}[b!] 
\centerline{\epsfig{file=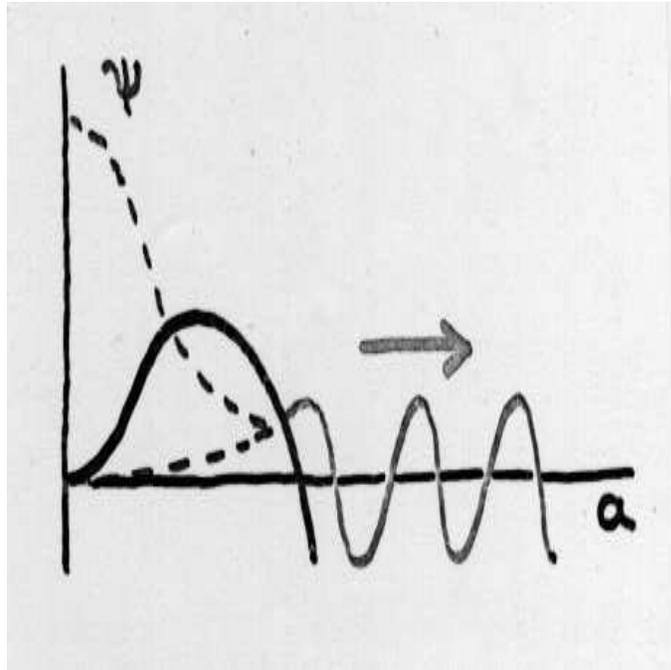,height=3.5in,width=3.5in}}
\vspace{10pt}
\caption{The tunneling wave function for the simple model (\ref{WDW}).}
\label{fig2}
\end{figure}


In a more realistic model, the constant vacuum energy density $\rho_v$
is replaced by the potential $V(\phi)$ of some scalar field $\phi$.
If $V(\phi)$ is a sufficiently slowly-varying function of $\phi$, one
finds the same result as before, with the replacement $\rho_v\to
V(\phi)$,
\beq
{\cal P}\sim\exp\left(-{3\over{8V(\phi)}}\right).
\label{prob}
\eeq
Eq.(\ref{prob}) can be interpreted as the probability distribution for the
initial values of $\phi$ in the ensemble of nucleated universes.  
The highest probability is obtained
for the largest values of $V(\phi)$ (and smallest initial size
$a_0$).  Thus, the tunneling wave function `predicts' that the
universe is most likely to nucleate with the largest possible vacuum
energy.  This is just the right initial condition for inflation.

In the general case, the wave function of the universe is defined on
superspace, which is the space of all 3-dimensional geometries and
matter field configurations,
$\psi [g_{ij}({\bf x}), \phi ({\bf x})]$,
where $g_{ij}$ is the 3-metric, and matter fields are represented by a
single field $\phi$.  The tunneling boundary condition can be extended
to full superspace by requiring that $\psi$ should include only
outgoing waves at the boundary of superspace, except the part of the
boundary corresponding to vanishing 3-geometries.

Alternatively, the tunneling wave function can be defined as a path
integral
\beq
\psi_T(g,\phi)=\int_\emptyset^{(g,\phi)}e^{iS},
\label{psiT}
\eeq
where the integration is over paths interpolating between a vanishing
3-geometry $\emptyset$ (`nothing') and $(g,\phi)$.  In other words,
the integration is over compact Lorentzian geometries
bounded by the 3-geometry $g$ with the field configuration $\phi$.  

At present these general definitions of the tunneling wave function
remain largely formal since we do not know how to solve the
Wheeler-DeWitt equation or how to calculate the path integral
(\ref{psiT}), except for simple models and small perturbations about them.

\section*{The Hartle-Hawking wave function}

The Hartle-Hawking wave function is expressed as a path integral over
compact Euclidean grometries
bounded by a given 3-geometry $g$,
\beq
\psi_{HH}(g,\phi)=\int^{(g,\phi)}e^{-S_E}.
\label{HH}
\eeq
The Euclidean rotation of the time axis,
$t\to i\tau$,
is often used in quantum field theory because it improves the
convergence of the path integrals.  However, in quantum gravity the
situation is the opposite.  The gravitational part of the Euclidean
action $S_E$ is unbounded from below, and the integral (\ref{HH}) is
badly divergent.  One can hope that the problem will somehow be fixed
in the future theory of quantum gravity, but at present we cannot
meaningfully define an integral such as (\ref{HH}).

In practice, one assumes that the dominant contribution to the path
integral is given by the stationary points of the action (the
instantons) and evaluates $\psi_{HH}$ simply as
$\psi_{HH}\sim e^{-S_E}$.
For our simple model, $S_E\approx -3/8V(\phi)$ and
\beq
{\cal P}\sim\exp\left(+{3\over{8V(\phi)}}\right).
\label{probHH}
\eeq
The wave function $\psi_{HH}(a)$ for this model is shown in Fig.3.
It has only the growing solution under the barrier and a superposition
of ingoing and outgoing waves with equal amplitudes in the classically
allowed region.  This wave function appears to describe a contracting
and re-expanding universe.

\begin{figure}[b!] 
\centerline{\epsfig{file=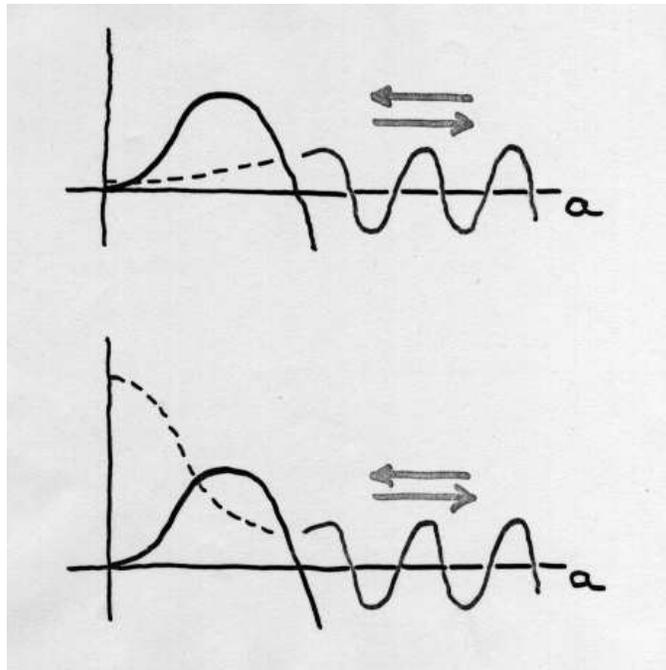,height=3.5in,width=3.5in}}
\vspace{10pt}
\caption{Hartle-Hawking (top) and Linde (bottom) wave functions for
the model (\ref{WDW}).}
\label{fig1}
\end{figure}

The distribution (\ref{probHH}) is similar to Eq.(\ref{prob}) for the 
tunneling wave function, but
there is a crucial difference in sign.  The distribution (\ref{probHH}) is
peaked at the smallest values of $V(\phi)$, and thus the
Hartle-Hawking wave function tends to predict initial conditions that
disfavor inflation.  

In fairness, one has to admit that this objection is not fatal.
Without inflation, a tiny nucleated universe will never reach a
macroscopic size and will not evolve any living creatures, so there
will be nobody there to observe it.  If universes are predominantly of
this kind, then most of them will never be observed.  The right
question to ask, then, is not what a typical universe looks like, but
what a typical observer will see \cite{M}.  
The larger the universe is, the
more stars it contains, the more civilizations are likely to develop.
If we are a typical civilization, then we can expect
to live in a large and populous universe characterized by a large
amount of inflation \cite{M,LGB,Page}.
Moreover, in many models inflation is eternal to the future, provided
that the initial value of $V(\phi)$ is sufficiently large \cite{V83,L2}.
An eternally inflating universe produces an infinite number of
observers, and thus we expect to find ourselves in such a universe
with a 100\% probability, even if the probability of its nucleation is
very low.  

\section*{The Linde wave function}

Linde suggested that the wave function of the universe is given by a
Euclidean path integral like (\ref{HH}), but with the Euclidean time rotation
performed in the opposite sense,
$t\to +i\tau$,
yielding
\beq
\psi_L=\int^{(g,\phi)}e^{+S_E}.
\eeq
For our simple model, this wave function gives the same nucleation
probability (\ref{prob}) as the tunneling wave function.

The problem with this proposal is that the Euclidean
action is also unbounded from above and, once again, 
the path integral is divergent.
This divergence is even more disastrous than in the Hartle-Hawking
case, because now all integrations over matter fields and over
inhomogeneous modes of the metric are divergent.  
It is not clear how Linde's proposal can be extended beyond the
simplest model.  This problem of Linde's wave function makes it an
easy target, and I suspect it is for this reason that Stephen likes to
confuse $\psi_L$ and $\psi_T$ and refer to both of them as ``the
tunneling wave function''.  In fact, the two wave functions are quite
different, even in the simple model (\ref{WDW}) \cite{discord}.  
The Linde wave
function includes only the decaying solution under the barrier and a
superposition of ingoing and outgoing modes with equal amplitudes
outside the barrier (see Fig. 3).

To summarize, the Hartle-Hawking and Linde wave functions have serious
problems with divergent integrals.  In addition, the Hartle-Hawking
wave function has a potential problem with inflation.  The tunneling
wave function appears to do reasonably well on both accounts.

\section*{Quantum creation of open universes}

\begin{figure}[b!] 
\centerline{\epsfig{file=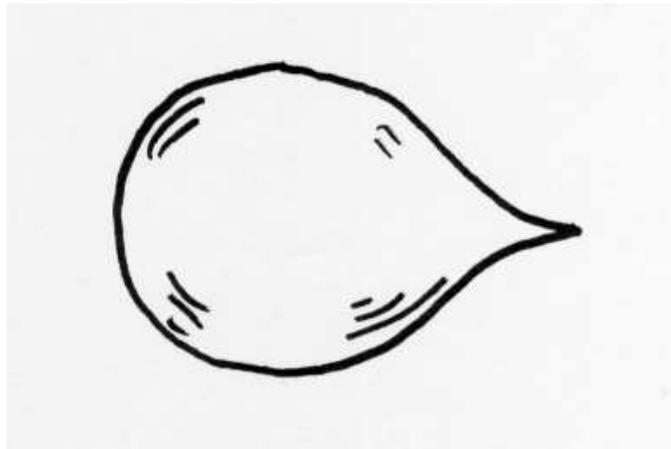,height=2.33in,width=3.5in}}
\vspace{10pt}
\caption{The Hawking-Turok instanton.}
\label{fig4}
\end{figure}

Hawking and Turok (HT) have recently argued \cite{19} that open
universes can be spontaneously created from nothing and suggested an
instanton to describe this process.  They considered a model with a
very simple potential, 
\begin{equation}
V(\phi) = {1\over{2}}m^2\phi^2.
\label{V}
\end{equation}
This model is known not to have regular
instanton solutions, and indeed, the HT instanton is
singular. Geometrically, it is like a sphere with a thorn, the tip of
the thorn being the singularity where the curvature and the scalar
field are infinite (see Fig. 4). HT point out, however, that the singularity is
integrable and the instanton action is finite. Analytic continuation
of this instanton gives a closed, singular spacetime.  A part of this
spacetime, is isometric to an open Robertson-Walker universe. The
singularity has the form of an expanding singular bubble. 
However, it never hits an observer in the Robertson-Walker
part of the universe, and HT argue that the singularity is therefore
not a problem \cite{24}.

HT instantons have a free parameter corresponding to the strength of
the singularity. As this parameter is varied, the density parameter
$\Omega$ of the open universe also changes, and HT use an anthropic
approach to find the most probable value of
$\Omega$.

I think there are serious problems with HT approach.  
In a singular instanton, the field equations are not
satisfied at the singularity, and such an instanton is not, therefore,
a stationary point of the action.  It is not clear why such instantons
should dominate the path integral.  Moreover, if HT instanton is
allowed, we will then have to admit a host of other instantons with
integrable singularities.  I am going to give some examples of such
instantons and argue that they lead to unacceptable consequences.

First, I will  construct a singular
instanton for nucleation of open universes which has a lower action than 
the Hawking-Turok one.  Take two copies of HT instanton, cut off their
thorns and match what remains of the instantons across the cut, as in Fig.5.  
\begin{figure}[b!] 
\centerline{\epsfig{file=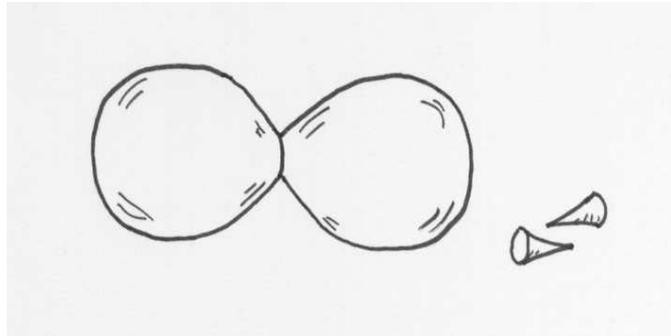,height=1.75in,width=3.5in}}
\vspace{10pt}
\caption{This singular instanton is made out of two HT instantons with
their thorns cut off.  It has a lower action than the HT instanton.}
\label{fig5}
\end{figure}
The resulting instanton
has an integrable domain-wall-type singularity at the matching
surface.\footnote{Note that the field $\phi$ should be continuous
across the matching surface in order for the action to be finite.}
Its action is about twice as negative as that of the HT instanton.  One can
go further and use HT-type instantons with thorns on both
sides\footnote{Instantons with two singular thorns 
have been discussed by
R. Bousso and A. Linde \cite{21}} to
construct singular instantons of arbitrarily large negative action
(see Fig.6).  In the HT approach, such instantons should completely
dominate the path integral.

\begin{figure}[b!] 
\centerline{\epsfig{file=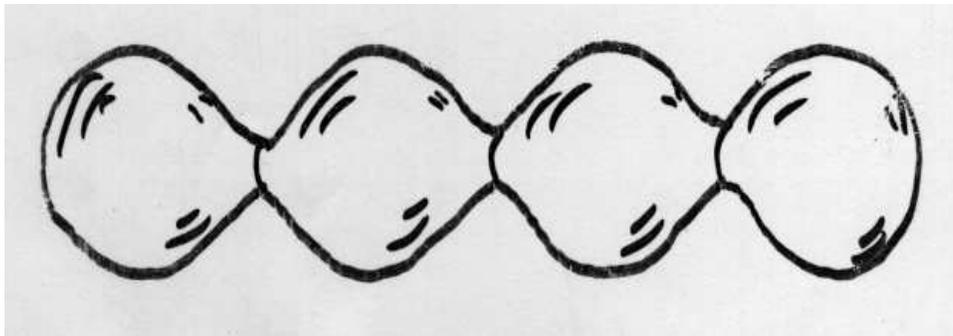,height=1.75in,width=5in}}
\vspace{10pt}
\caption{In this way one can construct singular instantons of
arbitrarily large negative action.}
\label{fig6}
\end{figure}

As a second example, for the same model (\ref{V}) I have constructed an
asymptotically-flat singular instanton \cite{23}. Geometrically, it
looks like a flat space with a thorn. The behavior of the fields near
the singularity is identical to that in the HT instanton and the
action is finite, so there is absolutely no reason to reject my
instanton if HT instanton is legitimate. The analytic continuation of
my instanton gives a flat space with a singular sphere which expands
at a speed close to the speed of light. If this were indeed a
legitimate instanton, then we would have to conclude that flat space
is unstable with respect to nucleation of singular bubbles. The
nucleation probability can be made very high by adjusting the strength
of the singularity, and since this strength is a free parameter, the
universe in this picture would have already been overrun by expanding
singular bubbles. Since this is in a glaring contradiction with
observations, we have to conclude that HT instanton, as it stands,
cannot be used to describe the creation of open universes.


Let me conclude by pointing out one thing that I think is good about
the HT instanton.  It has invigorated the debate about the basic
issues of quantum cosmology and will hopefully lead to a better
understanding of some of these issues.


\begin{references}
\bibitem{HH}
Hartle J.B. and Hawking S.W., Phys. Rev. {\bf D28}, 2960 (1983).

\bibitem{L1}
Linde A.D., Lett. Nuovo Cimento {\bf 39}, 401 (1984).

\bibitem{V84}
Vilenkin A., Phys. Rev. {\bf D30}, 509 (1984).

\bibitem{V86}
Vilenkin A., Phys. Rev. {\bf D33}, 3560 (1986).








\bibitem{M}
Vilenkin A., Phys. Rev. Lett. {\bf 74}, 846 (1995).

\bibitem{LGB}
Garcia-Bellido J. and Linde A.D., Phys. Rev. {\bf D51}, 429 (1995).

\bibitem{Page}
Page D.N., Phys. Rev. {\bf D56}, 2065 (1997).

\bibitem{V83}
Vilenkin A., Phys. Rev. {\bf D27}, 2848 (1983). 

\bibitem{L2}
Linde A.D., Phys. Lett. {\bf B175}, 395 (1986).

\bibitem{discord} Vilenkin, A., Phys. Rev. {\bf D58}, 067301 (1998).

\bibitem{19} Hawking S.W. and Turok N.G., Phys. Lett. {\bf B425}, 25
(1998).


\bibitem{24} Turok N.G. and Hawking S.W., Phys. Lett. {\bf B 432},
271 (1998).

\bibitem{23} Vilenkin A., Phys. Rev. {\bf D57}, 7069 (1998).

\bibitem{21} R. Bousso and A. D. Linde, Phys. Rev. {\bf D58}, 083503 (1998).


\end{references}
\end{document}